\documentclass[12pt]{article}
\usepackage[dvips]{graphicx}
\usepackage{amsfonts}

\title{Application of information and complexity theories to public opinion polls.
The case of Greece (2004-2007).}

\author{C.~P.~Panos\footnote{\texttt{e-mail:\,
        chpanos\,@\,auth.gr}}\,,
        K.~Ch.~Chatzisavvas\footnote{\texttt{e-mail:\,
        kchatz\,@\,auth.gr}}\,,
\\
 {\it  Department of Theoretical Physics,}\\
        {\it Aristotle University of Thessaloniki,}\\
                {\it  54124 Thessaloniki, Greece}
 }

\date{6 November 2007}

\begin{document}

\maketitle

\begin{abstract}

A general methodology to study public opinion inspired from
information and complexity theories is outlined. It is based on
probabilistic data extracted from opinion polls. It gives a
quantitative information-theoretic explanation of high job
approval of Greek Prime Minister Mr. Constantinos Karamanlis
(2004-2007), while the same time series of polls conducted by the
company Metron Analysis showed that his party New Democracy (abbr.
ND) was slightly higher than the opposition party of PASOK -party
leader Mr. George Papandreou. It is seen that the same
mathematical model applies to the case of the popularity of
President Clinton between January 1998 and February 1999,
according to a previous study, although the present work extends
the investigation to concepts as complexity and Fisher
information, quantifying the organization of public opinion data.

\end{abstract}

\section{Introduction}

Public opinion expressed in results of elections and opinion polls
have been studied widely using traditional statistics. An
alternative approach is information theory, which can be applied
to probabilistic data. Electoral data can be easily transformed
from percentages to probabilities. Thus the use of information
theory to investigate public opinion about political parties and
the conduct of governments and its opposition is obvious, but has
not been carried out so far. Such an application, the only one to
our knowledge, is an analytical approach to interpret the public's
high job approval rating for president Clinton \cite{Kulkarni99}.
This rating has been high and nearly constant between January 1998
and February 1999, despite the well known unfavorable conditions
for the US president in that period. Such a high rating could be
explained partially, but is still considered unusual for several
reasons \cite{Kulkarni99}. The political situation in Greece in
the recent three years is completely different than the United
States of the years 1998-1999. However, an interesting (parallel)
question arises: Greek Prime Minister Karamanlis enjoyed a high
job approval (2004-2007), although his party New Democracy
approval by the public was just $1\%$ higher than the opposition
party PASOK, headed by George Papandreou. We note that we used
statistical data from a specific Greek opinion polls company
(Metron Analysis) in the period 2004-2007, stopping just three
months before the latest parliament elections in Greece (September
2007). It is of interest to try to clarify the above striking fact
by extending the usual statistical treatment to Shannon's
information theory \cite{Shannon48}. Information theory was used
for the first time in telecommunications in the late 40's. Our aim
is to investigate the possibility to extract some general,
qualitative conclusions from typical opinion polls in Greece,
employing the tools of information and complexity theories. As we
mentioned above, our inspiration comes from a similar study in the
United States. Although the political systems and the conditions
in the USA and Greece are very different, our work leads to the
same mathematical model. It is seen that information-theoretic
methods can be used to extend the results of usual statistics,
which illuminate certain statistical data of public opinion.
Information theory can proceed further towards an interpretation,
in some sense, of statistical processes. The use of the logarithm
in the definition of information entropy smooths small differences
in statistical data from various companies and yields the same
qualitative conclusions. This illustrates the strength of
information theory to give quantitative (numerical) answers to
qualitative questions.

\section{Elements of Information Theory}

Specifically information entropy $S$, corresponding to a
probability distribution $\{p_{i}\}, i=1,2,\ldots,N$ of $N$ events
occuring with probabilities $p_{i}$, respectivelly, can be defined
as
\begin{equation}\label{eq:eq1}
    S=-\sum_{i=1}^{N} p_{i}\, \log{p_{i}}, \quad \mbox{where}\,\,
    \sum_{i=1}^{N} p_{i}=1 \,\,\mbox{(normalization)}
\end{equation}

$S$ is an information theoretic quantity which takes into account
all the moments of a probability distribution and can be
considered, in a sense, superior to traditional statistics
employing the well-known quantities of average value and variance.
$S$ in relation (\ref{eq:eq1}) is measured in bits (if the base of
logarithm is 2), nats--natural units of information (if the base
is e) and Hartleys (if the base is 10). In the present paper the
base is 10, for the sake of comparison with \cite{Kulkarni99}.
However, one case can be transformed to the other one, by
multiplying with just a constant.

Definition (\ref{eq:eq1}) represents the average information
content of an event, which occurs with a specific probability
distribution $\{p_{i}\}$. The use of the logarithm is justified
because in such a way $S$ obeys certain mathematical and intuitive
properties expected from a quantity related to information content
of a probability function. Specifically, $S$ is positive and the
joint information content of  two simultaneous independent events
translate to the addition of the corresponding information
measures of each event e.t.c. For more properties and a
pedagogical description see \cite{Shannon48}.

$S$ is maximum for an equiprobable or uniform probability
distribution
$p_{1},p_{2},\ldots,p_{N}=\displaystyle{\frac{1}{N}}$, i.e.
$S_{\rm max}=\log{N}$. $S$ is minimum when one of the $p_{i}$'s is
1 ($p_{i}=1$) and all the other $p_{i}$'s are 0, i.e. $S_{\rm
min}=0$, under the convention that $0\,\log{0}=0$. In this case,
one of the outcomes is certain, while all the other ones are
impossible to occur.

$S$ represents a measure of information content of a probabilistic
event, i.e. the average number of "Yes" or "No" questions needed
to specify the event (in the case of bits). $S$ is reciprocal to
the degree of surprise of an event, i.e. the least probable event
has the most information and vice versa.

We give a simple example in order to understand the meaning of
relation (\ref{eq:eq1}). Let us ask to a certain number of people
the following question: \emph{Is C. Karamanlis suitable for the
position of Prime Minister of Greece?} We receive answers with
percentages and corresponding probabilities $p_{1}$ (Yes), $p_{2}$
(No), $p_3$ (Something Else). A direct application of
(\ref{eq:eq1}) for the normalized probability distribution
$\{p_{1},p_{2},p_{3}\}$, where $p_{1}+p_{2}+p_{3}=1$ ($N=3$) gives
the information content in Hartleys of that set of probabilities.

In the case of a uniform (equiprobable) distribution i.e.
$p_{1}=p_{2}=p_{3}=1/3\simeq 33.3\%$, relation (\ref{eq:eq1})
gives $S=S_{\rm max}=0.4471$. This is the maximum information
entropy with uniform probability distribution ($N=3$). This can be
interpreted as a distribution of complete ignorance (unbiased) in
the sense that a specific answer does not contain more information
than any other one. A case of maximum entropy $S$ corresponds to a
minimum amount of information $I$ about our question. Thus
information $I$, is reciprocal with $S$

\begin{equation}\label{eq:eq2}
    I\sim \frac{1}{S}
\end{equation}

The above convention agrees with our intuition, i.e. the
information content of an event corresponding to a probability
distribution can be quantified by the magnitude of our surprise
after the event has occurred or how unpredictable is the outcome.

The case of equiprobable distribution for $N=2$, i.e.
$N_{1}=N_{2}=1/2=50\%$ occurred in the recent general parliament
elections in Italy (April 2006). There were two large coalition of
parties and one of the coalitions won with a slight difference in
votes, about 40,000 -while the number of votes was about
40,000,000. Thus, with real results $49.8\%$ versus $49.69\%$, we
can consider with a very satisfactory approximation that
$p_{1}\simeq p_{2}\simeq 0.5=50\%$. The application of information
theory in this case gives $S=S_{\rm max}=\log_{N}2=1$ bit (base of
the logarithm equals 2). This fact is completely equivalent with
throwing a fair coin (equal probability for the two results
heads-tails) or with the question Yes-No (equiprobable) which
coalition will win. That means that $S=S_{\rm max}$ gives
$I=I_{\rm min}$ and the minimum information can be interpreted as
a complete homogenization of the public opinion about the two
coalitions. In other words, the results of elections in Italy
correspond to the random throw of a fair coin i.e. a complete lack
of knowledge of the voters. Our observation does not intend to
depreciate the process of elections, the culmination of democracy,
but it is an extreme case with maximum possible information
entropy $S_{\rm max}$.

There are other measures of information such as Onicescu's
information energy $E$ \cite{Onicescu66} and Fisher's information
$F$ \cite{Fisher25}. Shannon's information is a global measure,
while Fisher's is a local one i.e. $S$ does not depend on the
ordering of the probabilities $\{p_{i}\}$, while $F$ does depend,
due to the existence of the derivative of the distribution in its
definition.

Their definitions are given below together with appropriate
comments. It is stressed that all are based on the same
probability distributions $\{p_{i}\}$ as $S$.

Landsberg's definition of disorder $\Delta$ \cite{Landsberg98} is
\begin{equation}\label{eq:3a}
    \Delta=\frac{S}{S_{\rm max}}
\end{equation}
and order
\begin{equation}\label{eq:3b}
    \Omega=1-\Delta=1-\frac{S}{S_{\rm max}}, \quad
    (\Delta+\Omega=1)
\end{equation}

Disorder $\Delta$ is a normalized disorder ($0<\Delta<1$).
$\Delta=0$ (zero disorder, $S=0$) corresponds to complete order
$\Omega=1$ and $\Delta=1$ (complete disorder, $S=S_{\rm max}$)
corresponds to zero order $\Omega=0$. $\Delta, \Omega$ enable us
to study the organization of data, described probabilistically.

The next important step is the \emph{statistical complexity}
\begin{equation}\label{eq:eq4}
    \Gamma_{\alpha,\beta}=\Delta^{\alpha} \Omega^{\beta}
\end{equation}
defined by Shiner-Davison-Landsberg (SDL) \cite{Shiner99}, where
$\alpha$ is the strength of disorder and $\beta$ is the strength
of order. In the present work we consider the simple case
$\alpha=1$ and $\beta=1$.

Another measure of \emph{complexity} is
\begin{equation}\label{eq:eq5}
    C=S D
\end{equation}
according to Lopez Ruiz-Mancini-Calbet (LMC) \cite{Lopez95}. Here
$D$ is the so-called \emph{disequilibrium} (or distance from
equilibrium) defined as
\begin{equation}\label{eq:eq6}
    D=\sum_{i=1}^{N} \left(p_{i}-\frac{1}{N}\right)^2
\end{equation}

SDL complexity $\Gamma_{\alpha,\beta}$ describes correctly the two
extreme cases of complete order and complete disorder, where we
expect intuitively zero complexity or organization of the data. An
example taken from the physical world is illuminating. A perfect
crystal (complete order) has $\Gamma=0$ and the same holds for a
gas (complete disorder) where $\Gamma=0$ as well. Thus (perfect)
crystals and gases are not interesting, lacking complexity or
organization. This is given by $\Gamma_{\alpha,\beta}$ and agrees
with intuition. Instead, for the information entropy we have $S=0$
for crystals and $S=S_{\rm max}$ for gases, which is not
satisfactory.

Thus extending from physics, $\Delta$, $\Omega$,
$\Gamma_{\alpha,\beta}$ and $C$ enable us to study quantitatively
the (organized) complexity of probabilistic data of opinion polls
and elections.

An other very important information measure is Fisher information
$F$ \cite{Fisher25}. Recently, there is a revival of interest for
Fisher information, culminating in two books \cite{Frieden04} and
\cite{Frieden06}, defined as

\begin{equation}\label{eq:eq8}
    F=\int \frac{\left(\frac{d\rho(x)}{dx}\right)^2}{\rho(x)}\,dx
\end{equation}
for a continuous probability distribution $\rho(x)$, which is
modified accordingly in the present work for discrete probability
distributions. Specifically, for a discrete probability
distribution $\{p_{i}\}$ employed in the present work, relation
(\ref{eq:eq8}) becomes

\begin{equation}\label{eq:eq8qa}
    F=\sum_{i=1}^{3} \frac{(p_{i+1}-p_{i})^2}{p_{i}}
\end{equation}

Thus the treatment of high job approval of Clinton in
\cite{Kulkarni99}, will be repeated for the case of the Greek
Prime Minister Constantinos Karamanlis and the Greek political
scene in the recent three years (2004-2007) and extended in the
present paper using new quantities e.g. $\Delta$, $\Omega$,
$\Gamma_{\alpha,\beta}$, $C$ and $F$ as functions of time.

\section{Results and Discussion}

We used statistical data for the public opinion coming from the
Greek opinion polls company \emph{Metron
Analysis}\footnote{http://www.metronanalysis.gr}. Specifically, we
focused our interest on the following three questions, presented
in Table \ref{tab:tab1}.

\begin{table}[h!]\label{tab:tab1}
\begin{tabular}{|r|p{11.0cm}|}
  \hline
  % after \\: \hline or \cline{col1-col2} \cline{col3-col4} ...
    Question A & \emph{Choose a political leader, who is, according to your
    opinion, most suitable for the position of Prime Minister of
    Greece.} \\
    \hline
    Answers & Karamanlis ($p_{1}$), Papandreou ($p_{2}$), and
    Other ($p_{3}$). \\
    \hline \hline
    Question B & \emph{Are you satisfied from Mr. Karamanlis or Mr. Papandreou as
    political leaders?} \\
    \hline
    Answers & from Karamanlis ($p_{1}$), from Papandreou ($p_{2}$), and
    from No one ($p_{3}$). \\
    \hline \hline
    Question C & \emph{Which party you wish to vote for?} \\
    \hline
    Answers & New Democracy (abbr. ND) -party leader Karamanlis ($p_{1}$),
    PASOK -party leader Papandreou ($p_{2}$), and Other party ($p_{3}$). \\
    \hline
\end{tabular}
    \caption{Questions A, B and C.}\label{tab:tab1}
\end{table}

\begin{figure}
    \centering
 \begin{tabular}{cc}
 \includegraphics[height=6.0cm,width=7.0cm]{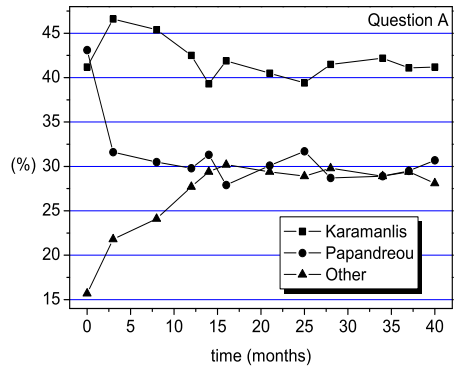}
 \\
 \includegraphics[height=6.0cm,width=7.0cm]{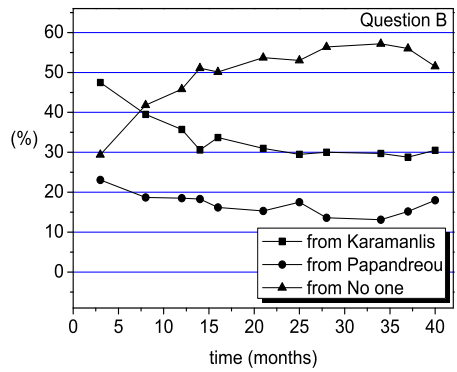}
 \\
 \includegraphics[height=6.0cm,width=7.0cm]{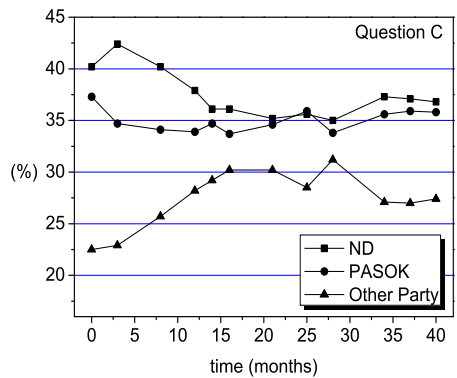}
 \end{tabular}
 \caption{Results for opinion polls for Questions A, B, and C
 respectively. The horizontal axes represent time in months
 ($t=0$ is March 2004), while the vertical axes show the
 corresponding percentages (Metron Analysis-Greece).}\label{fig:fig1}
\end{figure}

The corresponding answers (as probabilities), asked to a careful
chosen number of voters, are shown in the vertical axes of Fig.
\ref{fig:fig1} (for Questions A, B and C, respectively).

In all the figures the horizontal axis represents the time in
months, starting from $t=0$ (March 2004) to $t=40$ (June 2007).
Time $t=0$ is just before the parliament elections of April 2004
(winner ND) and time $t=40$ is 3 months before the latest
elections that took place in September 2007, with another victory
for ND.

\begin{figure}
 \begin{tabular}{cc}
 \includegraphics[height=6.0cm,width=7.0cm]{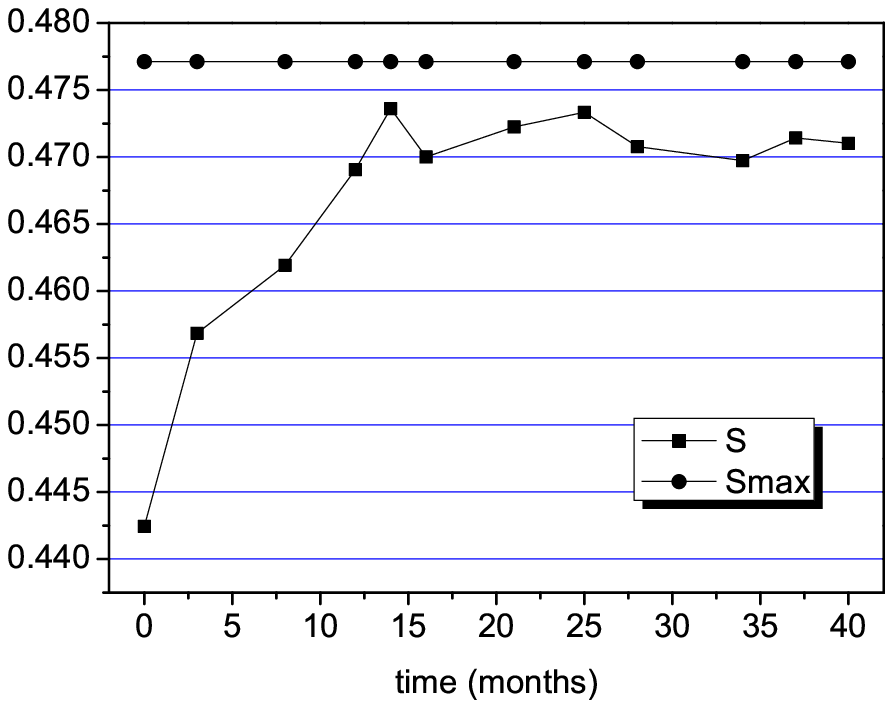}
 &
 \includegraphics[height=6.0cm,width=7.0cm]{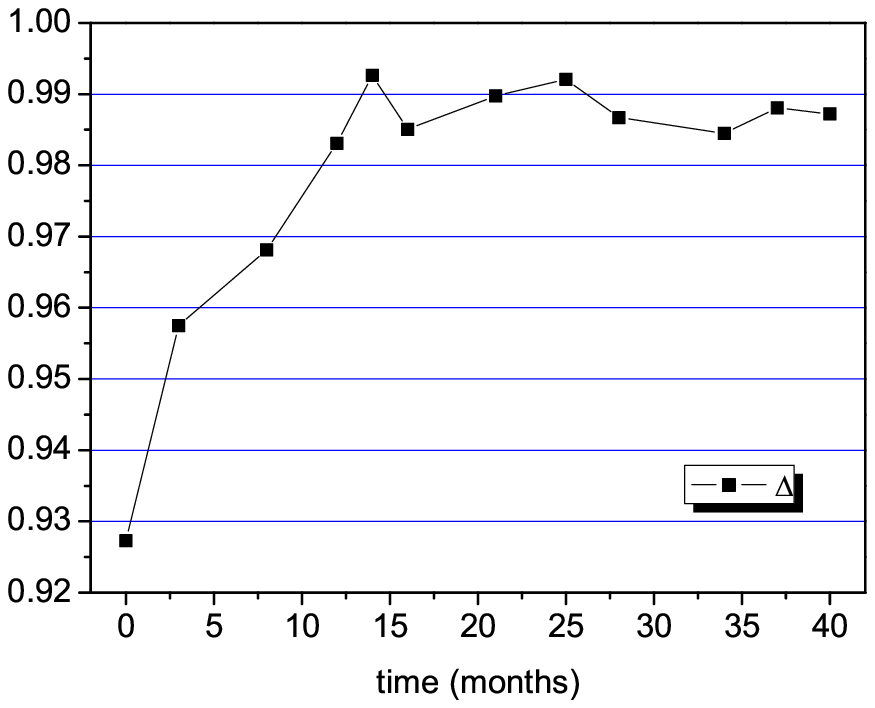}
 \\
 \includegraphics[height=6.0cm,width=7.0cm]{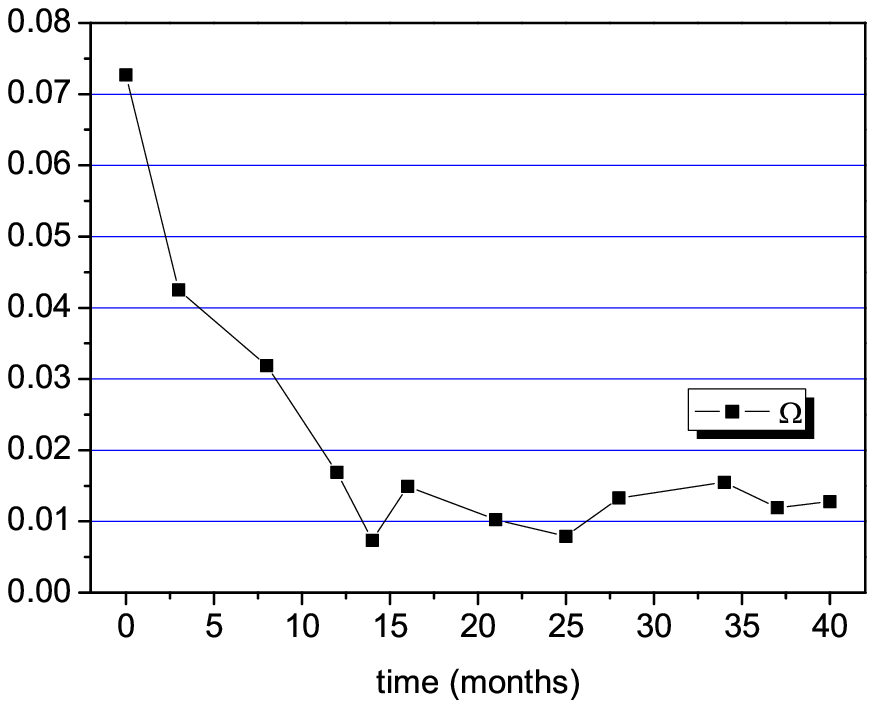}
 &
 \includegraphics[height=6.0cm,width=7.0cm]{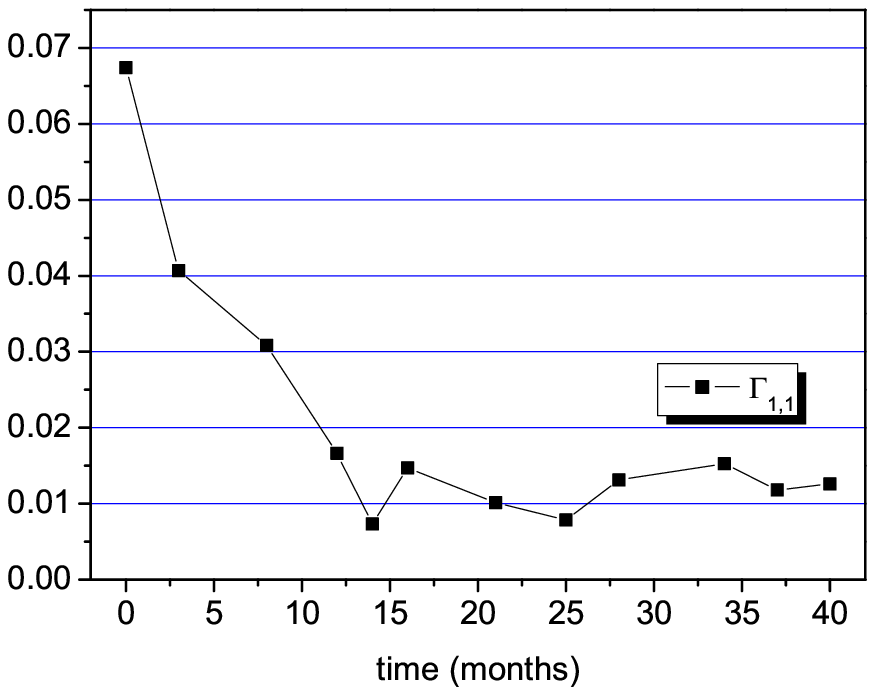}
 \\
 \includegraphics[height=6.0cm,width=7.0cm]{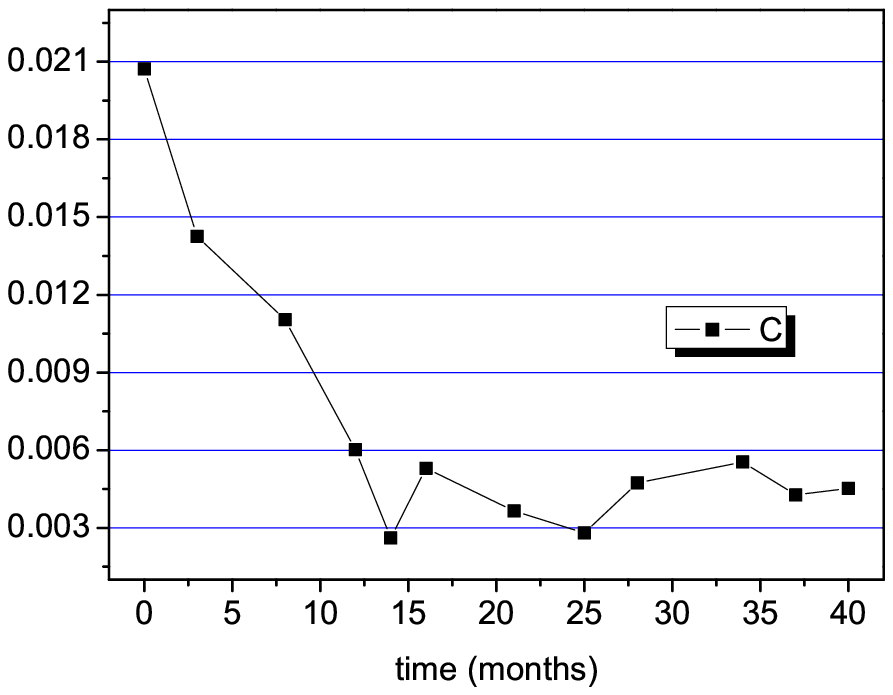}
 &
 \includegraphics[height=6.0cm,width=7.0cm]{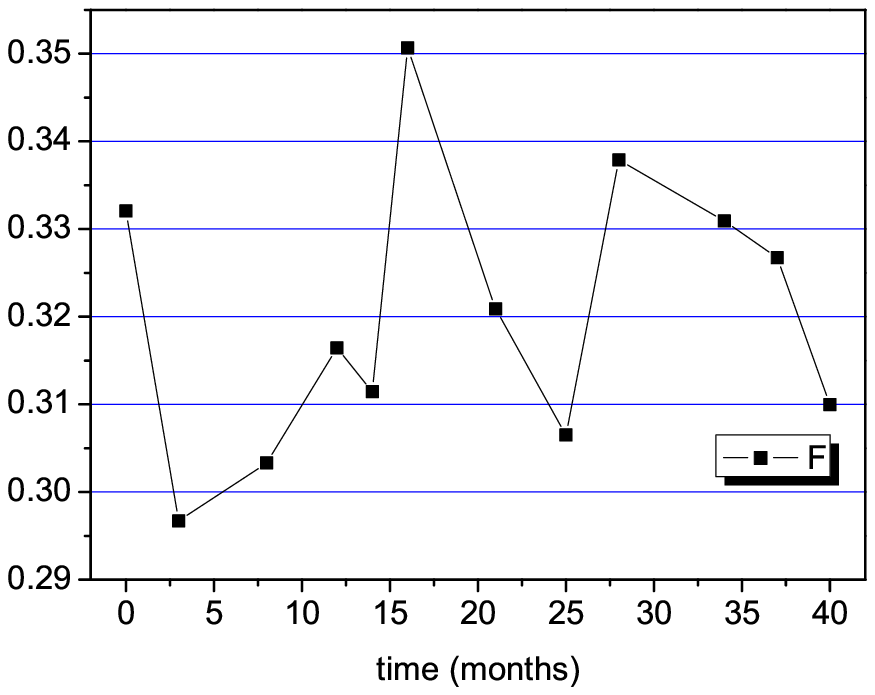}
 \end{tabular}
 \caption{Entropies $S^{(A)}(t)$ and $S^{(A)}_{\rm max}(t)$,
 Disorder $\Delta^{(A)}(t)$, Order $\Omega^{(A)}(t)$, Statistical Complexity
 $\Gamma^{(A)}_{1,1}(t)$, Complexity $C^{(A)}(t)$, and Fisher
 Information $F^{(A)}(t)$, for Question A.}\label{fig:fig2}
\end{figure}

Thus we have three sets of probabilities $\{p_{1},p_{2},p_{3}\}$
corresponding to Questions A, B, and C. In Fig. \ref{fig:fig2} we
present $S^{(A)}(t)$ and $S_{\rm max}(t)$, calculated using
(\ref{eq:eq1}), from probabilities of Fig. \ref{fig:fig1}
(Question A). We also present disorder $\Delta^{(A)}(t)$ and order
$\Omega^{(A)}(t)$, calculated from relations (\ref{eq:3a}) and
(\ref{eq:3b}). Statistical complexity $\Gamma^{(A)}_{1,1}(t)$,
complexity $C^{(A)}(t)$ and Fisher information $F^{(A)}(t)$ are
calculated employing relations (\ref{eq:eq4}), (\ref{eq:eq5}) and
(\ref{eq:eq8}) and are displayed in Fig. \ref{fig:fig2} as well.

\begin{figure}
 \begin{tabular}{cc}
 \includegraphics[height=6.0cm,width=7.0cm]{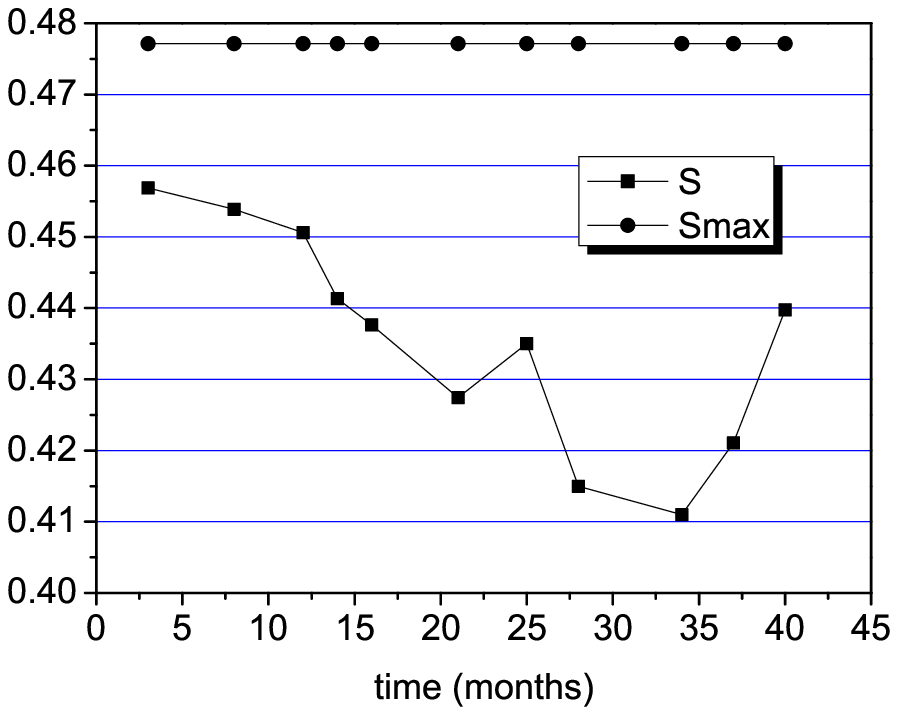}
 &
 \includegraphics[height=6.0cm,width=7.0cm]{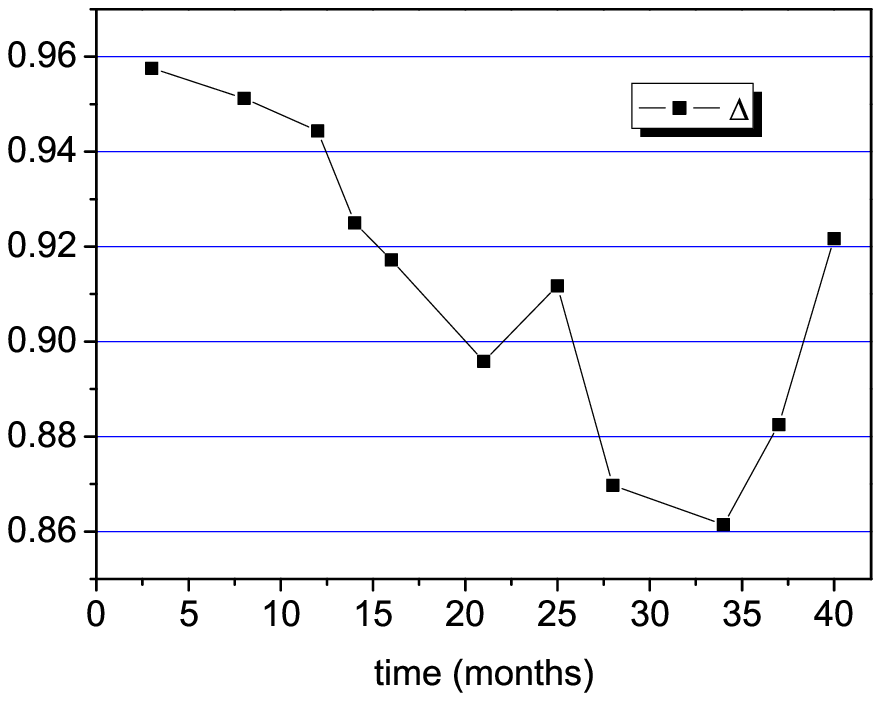}
 \\
 \includegraphics[height=6.0cm,width=7.0cm]{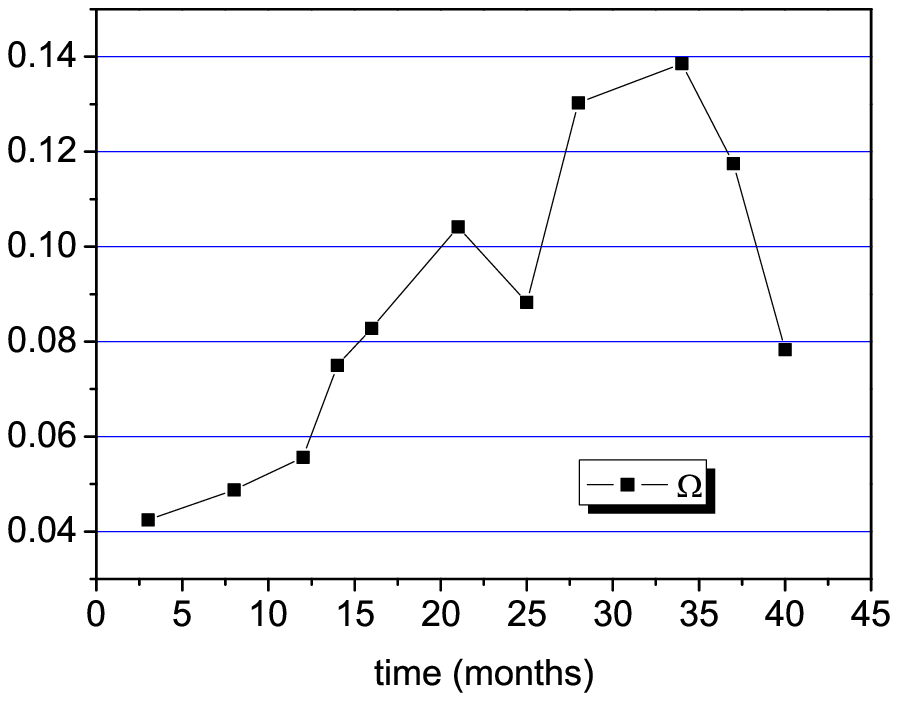}
 &
 \includegraphics[height=6.0cm,width=7.0cm]{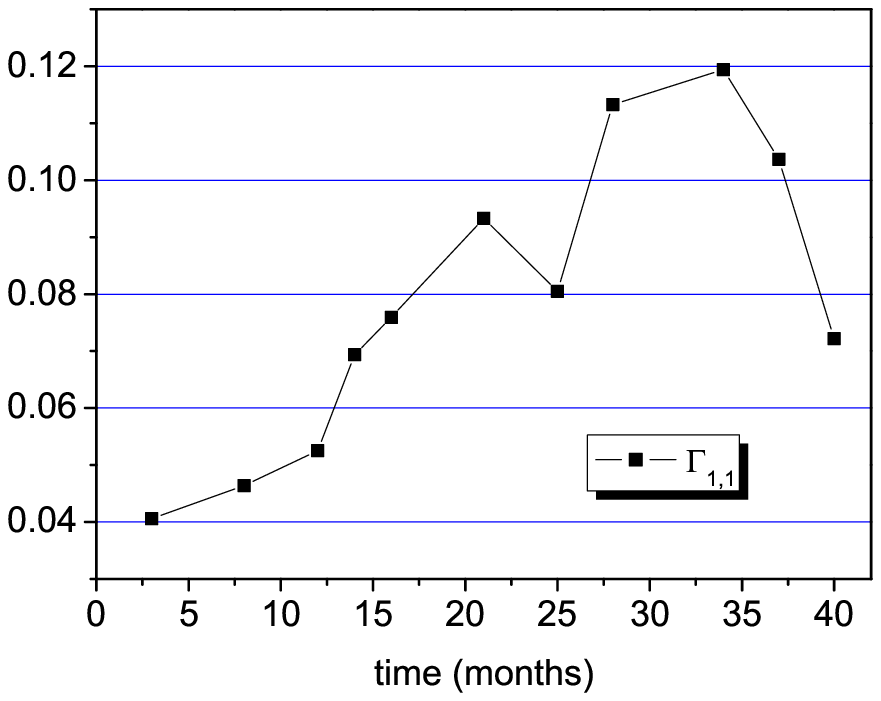}
 \\
 \includegraphics[height=6.0cm,width=7.0cm]{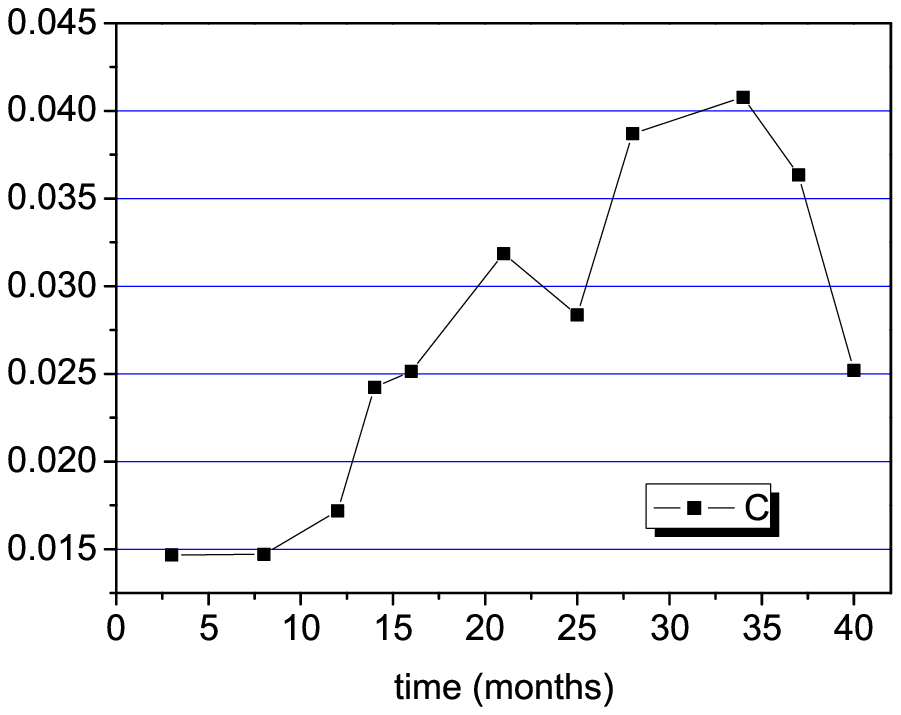}
 &
 \includegraphics[height=6.0cm,width=7.0cm]{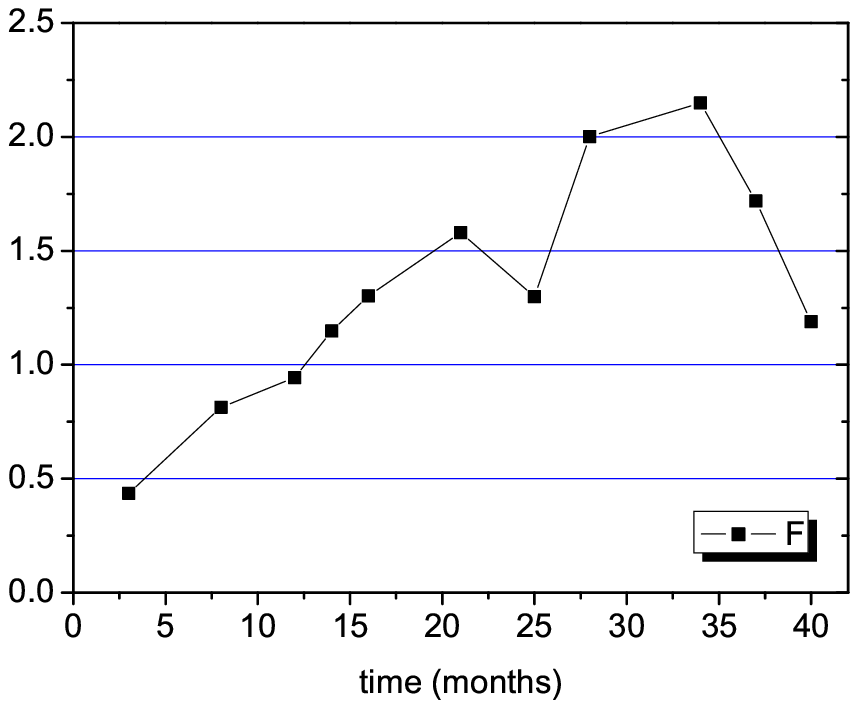}
 \end{tabular}
 \caption{Entropies $S^{(B)}(t)$ and $S^{(B)}_{\rm max}(t)$,
 Disorder $\Delta^{(B)}(t)$, Order $\Omega^{(B)}(t)$, Statistical Complexity
 $\Gamma^{(B)}_{1,1}(t)$, Complexity $C^{(B)}(t)$, and Fisher
 Information $F^{(B)}(t)$, for Question B.}\label{fig:fig3}
\end{figure}

\begin{figure}
 \begin{tabular}{cc}
 \includegraphics[height=6.0cm,width=7.0cm]{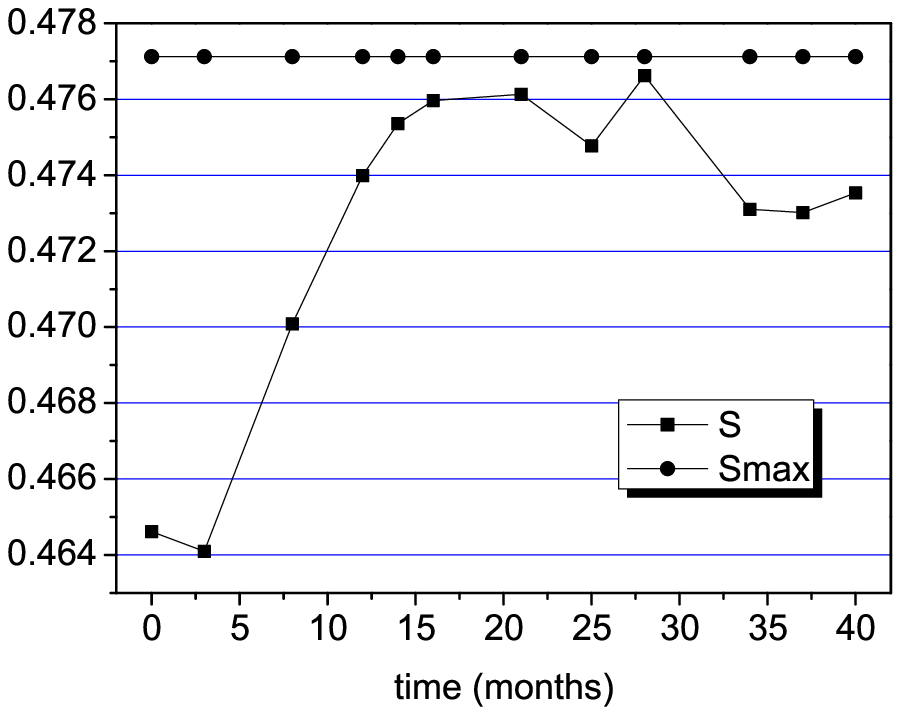}
 &
 \includegraphics[height=6.0cm,width=7.0cm]{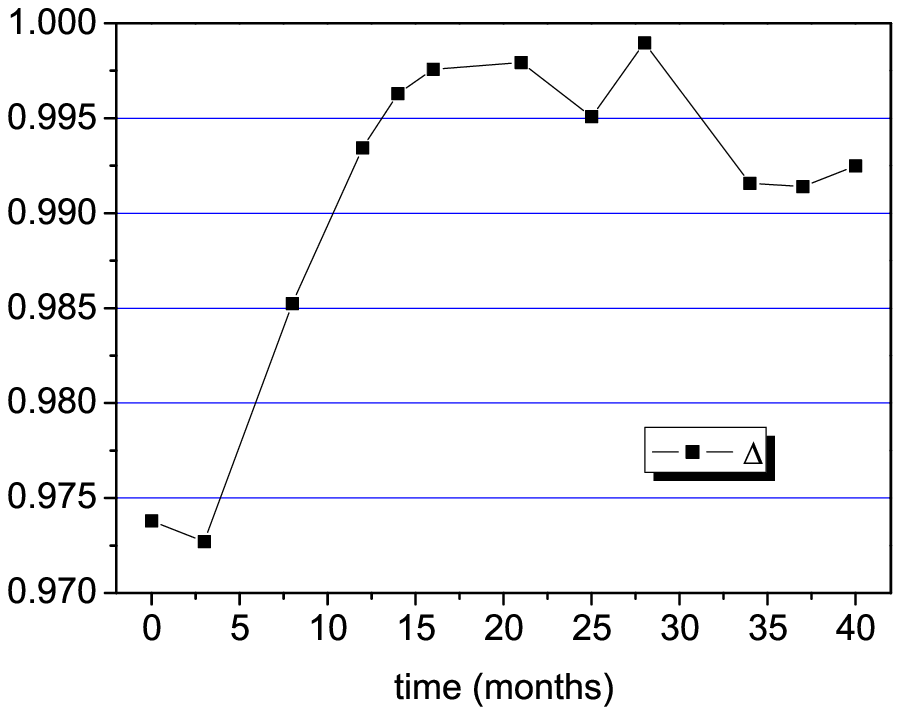}
 \\
 \includegraphics[height=6.0cm,width=7.0cm]{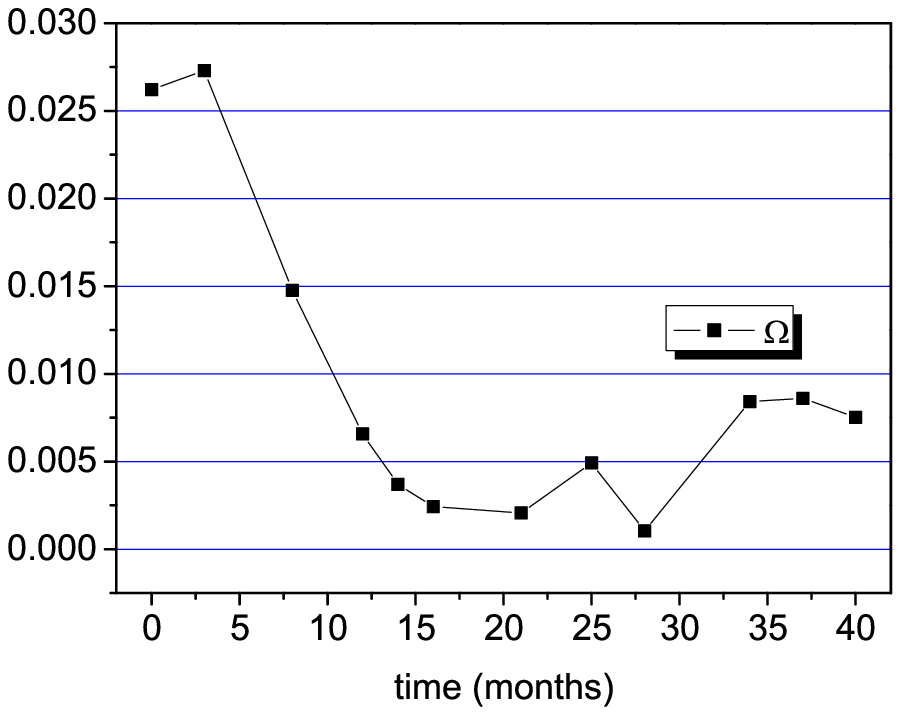}
 &
 \includegraphics[height=6.0cm,width=7.0cm]{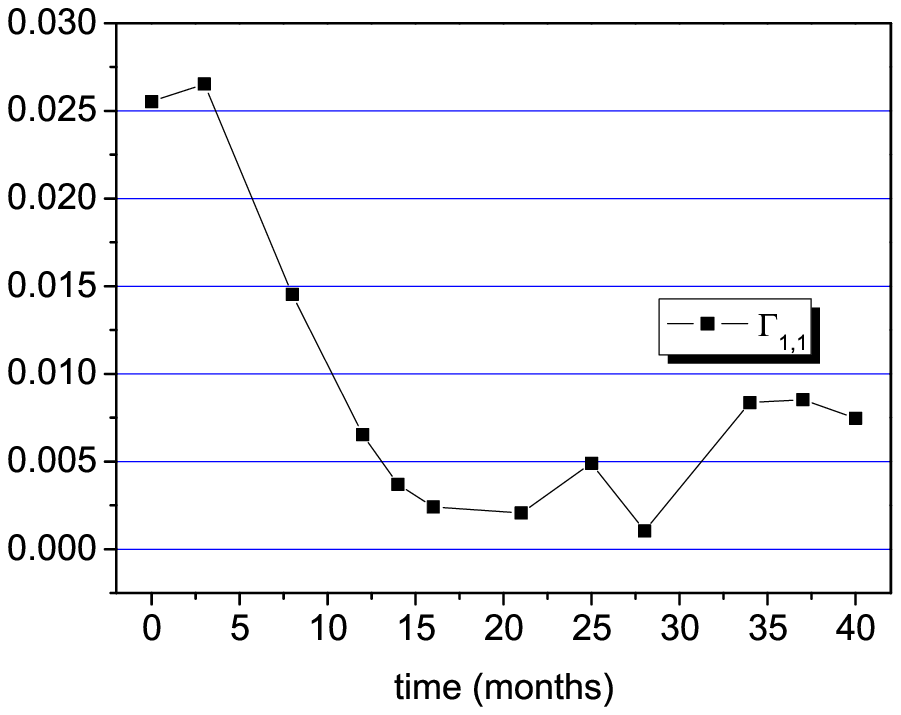}
 \\
 \includegraphics[height=6.0cm,width=7.0cm]{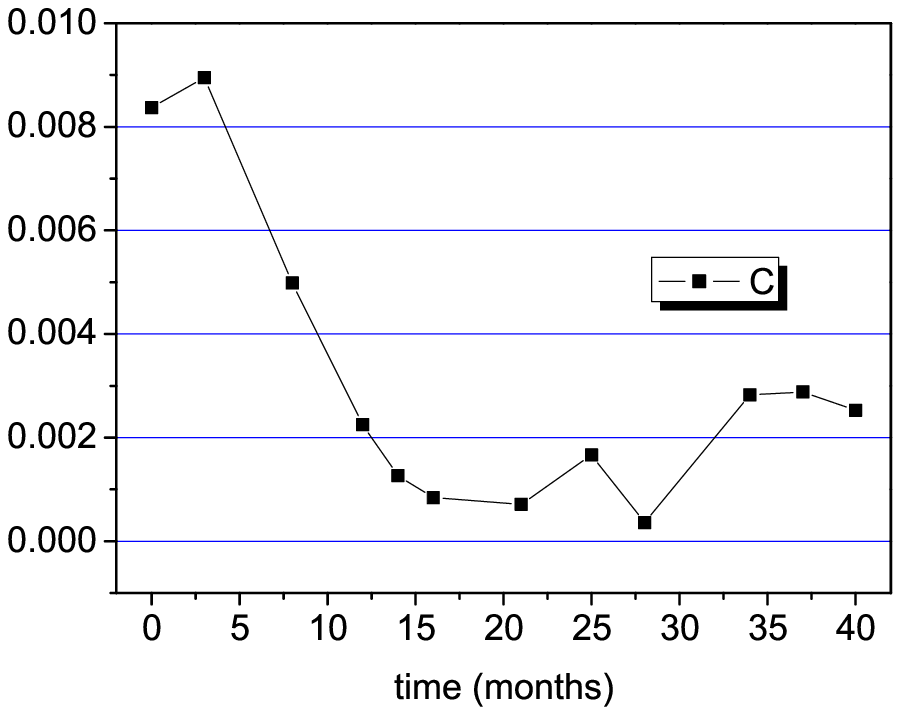}
 &
 \includegraphics[height=6.0cm,width=7.0cm]{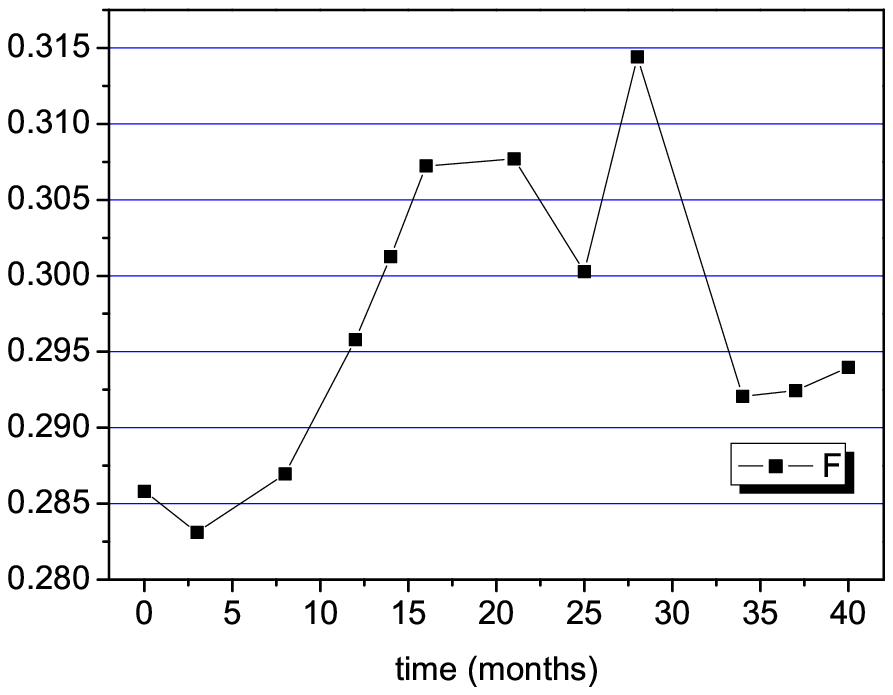}
 \end{tabular}
 \caption{Entropies $S^{(C)}(t)$ and $S^{(C)}_{\rm max}(t)$,
 Disorder $\Delta^{(C)}(t)$, Order $\Omega^{(C)}(t)$, Statistical Complexity
 $\Gamma^{(C)}_{1,1}(t)$, Complexity $C^{(C)}(t)$, and Fisher
 Information $F^{(C)}(t)$, for Question C.}\label{fig:fig4}
\end{figure}

In Fig. \ref{fig:fig3} we present $S^{(B)}(t)$, $\Delta^{(B)}(t)$,
$\Omega^{(B)}(t)$, $\Gamma^{(B)}_{1,1}(t)$, $C^{(B)}(t)$ and
$F^{(B)}(t)$ for Question B, while in Fig. \ref{fig:fig4} we
present $S^{(C)}(t)$, $\Delta^{(C)}(t)$, $\Omega^{(C)}(t)$,
$\Gamma^{(C)}(t)$, $C^{(C)}(t)$ and $F^{(C)}(t)$ for Question C.

In Question A, the entropy $S^{(A)}(t)$ shows an overall increase
as function of time and tends to the limit value of
$S_{max}=\log_{10}3=0.4771$ Hartleys. Taking into account that the
information $I$ is reciprocal to the entropy $S$ (\ref{eq:eq2}),
it is seen that the information possessed by the body of voters on
how suitable is Mr. Karamanlis, decreases as a function of time.
In a sense, Mr. Karamanlis achievement was to
\emph{compartmentalize} and make \emph{strongly independent}, the
opinion of voters for him as a person (high percentages in
Question A) compared to their opinion about his party (slightly
higher percentage of ND compared with PASOK). The results of the
parliament elections in September 2007 (ND-$42\%$, PASOK-$38\%$)
confirmed the idea of the Mr. Karamanlis' favorable and dominant
profile, despite the major event of the forest fires and the
collateral losses during August 2007.

The same trend can be extracted by the other information-theoretic
measures with the probabilities of Question A, e.g. the disorder
$\Delta$ increases while the order $\Omega$ decreases. This is a
mark that the organization of the statistics of the public opinion
decreases.

In Question B entropy as a function of time decreases and there is
a global minimum for $t=34$ months (December 2006). After that
global minimum it slightly increases, but stays far away from the
maximum value. Respectively, order $\Omega$ increases and shows a
global maximum for the same time ($t=34$ months), while the
behavior of the other information-theoretic measures is analogous.

It is seen from the figures that $I(t)=1/S(t)$, $\Omega(t)$,
$\Gamma_{1,1}(t)$ and $C(t)$, although their mathematical
definitions are different, show the same trends as functions of
time: For Question A all of them decrease with time, while for
Question B all increase. The same trend holds for Fisher
information $F^{(B)}(t)$, which is reciprocal to $S^{(B)}(t)$ and
analogous with $I^{(B)}(t)$. It is noted that this is the case for
a simple Gaussian probability distribution, as seen in the
literature \cite{Frieden06}. On the contrary, there is a striking
similarity in the behavior of $F^{(A)}(t)$ and $F^{(C)}(t)$
compared with with $S^{(A)}(t)$ and $S^{(C)}(t)$ respectively.
This can be attributed to the special meaning of Fisher's
information which is a local measure of information as contrasted
to Shannon's global measure. Specifically, changes in $S^{(A)}(t)$
or $S^{(C)}(t)$ are amplified as seen in the corresponding plots.
This is very interesting but however, can be considered as a
preliminary result taking into account that the presence of the
derivative in the definition of $F$ cannot be reflected correctly
in a case of a small discrete set of probabilities ($N=3$).

Here we can make the following comment: If we switch from Question
A to Question B, the qualitative behavior of the results changes
drastically, it is almost inverse. So, in Question B the
information that is available to the public increases (while the
corresponding information in Question A decreases) and there is a
global maximum (minimum) in December 2006. Thus, we have, from a
mathematical (quantitative) point of view, a strong indication
that the opinion polls are seriously affected from the formulation
of the relevant questions.

The fact that Mr. Karamanlis has a favorable profile based on
Question A is escorted by a clearly less favorable profile
according to the results of Question B. Completely analogous is
the picture of Mr. Papandreou. The satisfaction rate is clearly
less than the suitability for Prime Minister rate. The survey
concerning president Clinton \cite{Kulkarni99} is based on a
classical question about job approval which is definitely more
objective (\emph{"Do you approve or disapprove of the way [name of
president] is handling his job as president?"}) and it has been
used to US polls since 1930's.

On the other hand, the question about the suitability of a party
leader for the position of the Prime Minister (it has been used in
Greek opinion polls, since the mid 1990's) is less objective. The
person who answers the question takes into account extra data not
related to a realistic assessment of the job and accomplishments
of a party leader. The answers can be affected by factors such as
the public image of the leader (the one that he shows and the one
that the media advertise). There is more space for expectation or
hope that in the future some issues are going to improve or are
going to work better, due to special characteristics of the
leader, his personality, his abilities etc. Question B about
satisfaction is more realistic but still is general and obscure.

Qualitatively, results for Question C are similar with the results
of Question A. An interesting remark concerning the available data
is that in the time interval between $t=12$ months (March 2005)
and $t=28$ months (June 2006), the entropy reaches the maximum
value and information practically  minimizes. This time interval
can be considered as the interim (meantime) between two elections.

A final (extreme) remark seems appropriate in order to demonstrate
the difference between the information theory point of view and
the classical statistics approach.

Suppose that the majority party achieves a crashing victory over
minority, e.g. $70\%$ over $30\%$. The entropy for this scenario
should be $S=0.611$ bits, while the corresponding maximum value is
$S_{\rm max}=0.693$, so entropy decreases (or equivalently
information increases) only by $12\%$. Thus, in terms of
information theory, a fact that is completely clear-cut from the
point of view of classical statistics, is not so important to give
a complete and fair assessment of the public opinion. This has
been outlined in \cite{Kulkarni99} as well.

It is obvious that information theory can serve as a useful tool
even for politics surveys, with more details than the present
work. Something that so far has not been done systematically.

\begin{flushleft}
\Large{\textbf{Acknowledgements}}
\end{flushleft}

The authors would like to thank Mr. Pantelis Savvidis (director of
the newspaper \emph{Macedonia} of Thessaloniki) and Mr. Dimitris
Katsantonis (executive of the \emph{Metron Analysis} for
Thessaloniki) for providing the data.

\clearpage
\newpage

%%%%%%%%%%%%%%%%%%%%%%%%%%%%%%%%%%%%%%%%%%%%%%%%%%%%%%%%

\clearpage
\newpage

%%%%%%%%%%%%%%%%%%%%%%%%%%%%%%%%%%%%%%%%%%%%%%%%%%%%%%%%

\clearpage
\newpage

%%%%%%%%%%%%%%%%%%%%%%%%%%%%%%%%%%%%%%%%%%%%%%%%%%%%%%%%%%%%%%%

\clearpage
\newpage

%%%%%%%%%%%%%%%%%%%%%%%%%%%%%%%%%%%%%%%%%%%%%%%%%%%%%%%%%%%%%

\clearpage
\newpage
%%%%%%%%%%%%%%%%%%%%%%%%%%%%%%%%%%%%%%%%%%%%%%%%%%%%%%%%%%%%%%%

%%%%%%%%%%%%%%%%%%%%%%%%%%%%%%%%%%%%%%%%%%%%%%%%%%%%%%%%%%%%%%%%%%%

\end{document}